\documentclass[twocolumn,amsmath,amssymb]{snp}
\usepackage{graphicx}% Include figure files
\usepackage{dcolumn}% Align table columns on decimal point
\usepackage{bm}% bold math
\begin{document}

\title{{\Large Rapidity distribution dependence of the transition energy in heavy-ion collisions}}% Force line breaks with \\

\author{\large Sanjeev Kumar$^1$}
\author{\large Suneel Kumar$^1$}
\email{suneel.kumar@thapar.edu}
\affiliation{$^1$School of Physics and Materials Science,
Thapar University, Patiala - 147004, (Punjab) INDIA}

\maketitle

\section*{Introduction}
The relation between the nuclear equation of state (EOS) and 
flow phenomena has been explored extensively in the simulations as well as by the experimentalist.
Recently the analysis of transverse-momentum dependence of elliptical flow has also been put forwarded\cite{Luka055, Andr055, Dani00}. The elliptical flow pattern of participant matter is affected by the presence of cold
spectators \cite{Dani00}. During the expansion, the particles emitted towards
the reaction plane can encounter the cold spectator pieces and, hence, get redirected. In
contrast, the particles emitted essentially perpendicular to the reaction plane are largely
unhindered by the spectators. Thus, for the beam energies leading to rapid expansion in the
vicinity of the spectators, elliptic flow directed out of the reaction plane (squeeze-out) is
expected. This squeeze-out is related with the pace at which expansion develops, and is,
therefore, related to the EOS. This contribution of the participant and spectator matter \cite{Dani00} in the intermediate 
energy heavy-ion collisions motivated 
us to perform a detailed analysis of the excitation function of elliptical flow over different 
regions of participant and spectator matter in term of rapidity distribution bins. Attempts shall also be made to parameterize the transition energy for the same. \\
\section*{The IQMD Model}
The model is modified version of QMD model\cite{Hart98}. In this model, the nucleons of target and projectile interact via two and three body Skyrme forces, Yukawa and Coulomb potential. A symmetry potential between protons and neutrons corresponding to 
Bethe-Weizsacker mass formula has been included. The detail of the model is discussed in ref.\cite{Hart98} by us and others.\\
\section*{Results and Discussion}
For the present analysis, simulations are carried out for thousand of events for the reaction of 
$_{79}Au^{197}~+~_{79}Au^{197}$ at semi-central geometry using a hard equation of state. The whole 
of the analysis is performed for light charged particles (LCP's)[1$\le$ A $\le$ 4].\\
\begin{figure}
\hspace{-1.75cm}\includegraphics[scale=0.9]{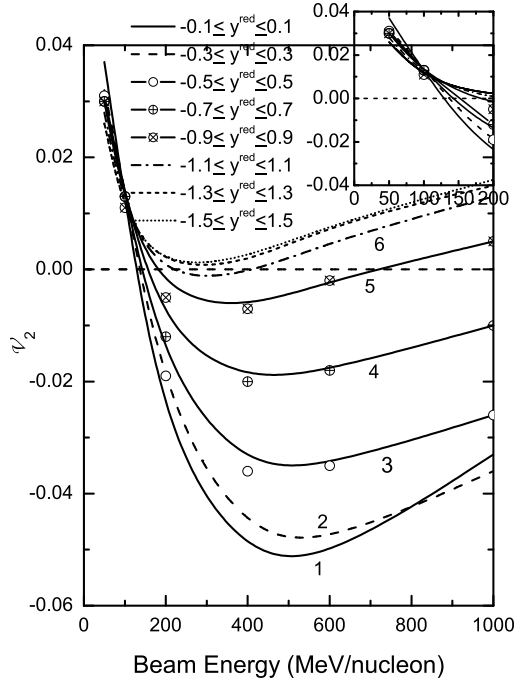}% Here is how to import EPS figure if you are using latex command to compile
\caption{\label{fig:1} The incident energy dependence of elliptical flow for LCP's collectively for 
projectile as well as target matter including mid-rapidity region. The different lines are at different size of the 
rapidity bin, which includes the participant as well as spectator matter.}
\end{figure}
\begin{figure}
\hspace{-1.75cm}\includegraphics[scale=0.8]{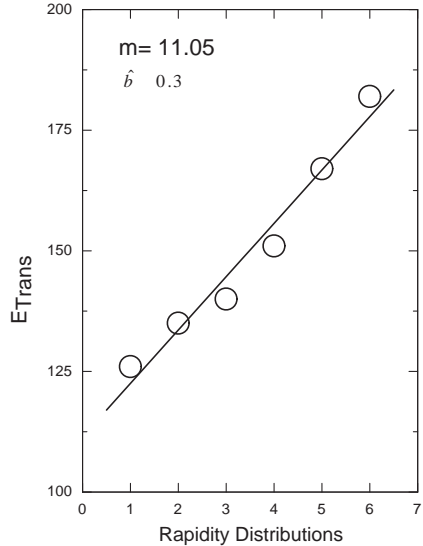}% Here is how to import EPS figure if you are using latex command to compile
\caption{\label{fig:2} The dependence of rapidity distribution on the transition energy. The figure is parameterized with the straight line interpolation $Y = mX~+~C$, where m is the 
slope.}
\end{figure}
In Fig.\ref{fig:1}, we present the excitation function of elliptical flow for the different conditions mentioned in the figure, where $Y_{c.m.}/Y_{beam}~=~Y^{red}$. The incident energy dependence of elliptical flow is well explained in the literature\cite{Luka055,Andr055}. On the other hand, the transition energy 
increases with rapidity region from $|Y^{red}|~\le~0.1$ and $|Y^{red}|~\le~1.5$. 
With the increase in the rapidity region, 
dominance of the spectator matter from projectile as well as target takes place that will further result in the dominance of 
the mean field up to higher energies. After the transition energy, the collective expansion is  found to have less squeeze
out with an increase in the rapidity region. As we know, the passing time for the spectator is very less compared to 
expansion time of the participant zone, leading to the decreasing effect of the spectator shadowing on the
participant zone. Due to this, the chances of the participant to move in-plane increases with increase in the rapidity bin. If one see carefully, no transition is observed after $|Y^{red}|~\le~1.1$. The inset in the figure shows interesting results: One can have a good 
study of elliptical flow below and above this particular incident energy with variation in the rapidity distribution.\\
Extracting the transition energy values from Fig.\ref{fig:1}, we have displayed the rapidity distribution dependence of transition energy in Fig.\ref{fig:2}.  The curve is fitted with straight line equation $Y~=~mX~+~C$, where m is slope of line and C is a constant. The transition energy is found to be sensitive
towards the different bins of rapidity distributions. It is observed that transition energy is found to increase
with the size of the rapidity bin for light charged particles.\\ 
In conclusion, the transition from in-plane to out-of-plane is observed only when the mid-rapidity region is included in the
rapidity bin otherwise no transition is observed. The transition energy is found to be strongly dependent on 
the size of the rapidity bin. The transition
energy is parameterized with a straight line interpolation. \\
%\section*{Acknowledgments}
%\acknowledgments
 %\end{acknowledgments}
%This work has been supported by the Grant no. 03(1062)06/ EMR-II, from %the Council of Scientific and
%Industrial Research (CSIR) New Delhi, Govt. of India.
%\ \\
%\noindent

\end{document}